# Percolative Model of Nanoscale Phase Separation in High Temperature Superconductors

J. C. Phillips

Bell Laboratories, Lucent Technologies (Retired),

Murray Hill, N. J. 07974-0636, USA

ABSTRACT

The nature of the phase diagrams of HTSC is clarified by discussing *two kinds* of phase diagrams, that of the host crystalline lattice, and that of the dopant glass. The latter is associated with changes in the electronic properties, while the former is much more accessible to direct experimental identification, by diffraction, of nanoscale phase separation. Careful examination of electronic properties in both the normal and superconductive states reveals that there are several electronic miscibility gaps in $YBa_2Cu_3O_x$ and $La_{2-x}Sr_xCuO_4$ that have been previously overlooked. Recent experiments on the pseudogap in $Bi_2Sr_{1.6}La_{0.4}CuO_y$ also reveal an electronic miscibility gap.

The phase diagrams of high temperature superconductors (HTSC) have been studied in great detail, and the effects of oxygen nonstoichiometry and lattice effects in $YBa_2Cu_3O_x$(YBCO) on phase transitions, structural distortions, and phase separation, have been reviewed (Kaldis 2001). The richness and complexity of these materials, as well as the nature of the underlying mechanisms, present many challenging and still unsolved problems. Here several trends in the data are summarized and explained, with emphasis on both the similarities and the differences between *two kinds* of phase diagrams: the crystalline host lattice structure, accessible to diffraction, and certain dopant electronic properties. The latter are the superconductive transition temperature $T_c$, and the pseudogap temperature $T^*$, where the planar resistivity becomes linear in T. Because the dopants do not have long-range order, correlations between electronic properties and the part of the host structure accessible to diffraction are not easily established.

Without doping, most HTSC are antiferromagnetic (AF) insulators, but with doping they become strange metals, with anomalous transport properties, radically different from those predicted by Fermi liquid theory (resistivities proportional to $T^2$). Does the doping merely create carriers in a real or virtual AF spin background, or does the doping alter the spatial lattice properties in a fundamental way? Generally the doping changes the host crystal structure from tetragonal to orthorhombic; in the case of YBCO, this change occurs at the metal-insulator transition at $x = 6.4$, where superconductivity also starts. In samples annealed under pressure above 240K, $dT_c/dP$ is by far the largest here (Fietz *et al.* 1996) among all the HTSC. Electron-spin interactions are insensitive to pressure, so that this suggests that it is orbital, not spin, properties that make HTSC possible.

For larger values of x in $YBa_2Cu_3O_x$, at least five different superlattice phases have been proposed based on $2a_0$ and $3a_0$ (and possibly $5a_0$ and $8a_0$) supercells normal to b-axis chains (Kaldis 2001, Andersen *et al.* 1999). However, most of these phases, together with wide *host* superlattice miscibility gaps, seem to have little effect on the electronic properties. The exception is the narrow *dopant* miscibility gap near $x = 6.75$, that has a very large effect on $T_c$ and on its derivatives with respect to pressure and uniaxial strain, especially along the a axis (Kraut *et al.* 1993). The effect of this gap is shown in Fig. 1 for two sets of samples from two recent experiments. The first set was powders multiply annealed in flowing oxygen, that gives a wide, flat 60K plateau, and a very narrow immiscibility dome (Akoshima and Koike 1998). The second set was single-crystal samples severely stressed by detwinning, that gives little plateau and a much wider dome (Segawa and Ando 2001). In YBCO the overdoped region occurs above $x = 6.94$, and is not easily studied, partly because it is so narrow (Kaldis 2001).

Overall there are many correlations between the host phase diagram and the dopant (electronic properties) phase diagrams. At the same time the coincidences appear to have systematic offsets, with the electronic miscibility gaps occurring at somewhat larger values of x, where one expects to find some excess oxygen dopants occupying nominally "empty" chain sites (Kaldis 2001). These excess dopants could contribute decisively to the electronic properties by acting as disordered orbital bridges between ordered chain segments of filamentary current paths. Thus, although the chains provide very effective orbital paths for orbital currents in YBCO, the well-correlated *host* and *dopant* aspects of the phase diagrams differ significantly, because the host structure is ordered, while the dopant structure is disordered.

It should also be noted that neutron and infrared studies of the vibronic spectra of YBCO, with both variable O doping and with Zn doping, have shown large changes with doping. These changes cannot be accounted for solely in terms of classical vibrational shell models with changes in host site occupancies (Reichardt *et al.* 1989; Homes *et al.* 1995; Phillips and Jung 2001). This indicates that the interactions between the host and the dopants are large. The interactions lead to partial (short-range) dopant ordering, and explain the similarities observed. It also suggests that dielectric screening by dopant current fluctuations can substantially affect the host structure, in a way not describable in terms of classical short-range interatomic interactions. Once again, the influence of spins is imperceptible.

These YBCO ideas can be applied to other HTSC phase diagrams. The most studied case is $La_{2-x}Sr_xCuO_4$ (LSCO), where the dopant is low-mobility Sr, not high-mobility O, and there are no simple native structural elements, like chains, that could be part of orbital current paths. However, LSCO is favorable in one very important respect, and that is that the entire phase diagram can be studied, from AF insulator, through HTSC intermediate phase, to overdoped Fermi liquid. We can now ask the following question: are there electronic miscibility gaps in the disordered, preparation-sensitive dopant part of the phase diagram, and are these separate and distinct from any crystalline host miscibility gaps or secondary phases?

The tetragonal-orthorhombic theme that dominates the YBCO phase diagrams through b-axis chain ordering is present also in LSCO. The optimal doping composition $x_0 = 0.16$ separates the overdoped range where the local (dopant) orthorhombicity is different from the macroscopic (host) one, from the underdoped range where the two agree (Haskel *et al.* 1996). (Thus in the optimally doped range there are disordered othorhombic nanodomains. The host orthorhombicity vanishes above $x = 0.21$, in the Fermi liquid, while nanodomain orthorhombicity persists.)

To deal with the problem of nanoscale percolative phase separation, one can use the following picture (see Fig. 2). One compares the phase diagram (Takagi *et al.* 1992) for the filling factor f(x), as measured either from the Meissner effect, or from the specific heat jump $\Delta C_p$ at $T_c$, on samples that have been annealed for long times at high temperatures, with $T_c(x)$. (Note that the filling factor $f(x)$ is the critical variable in any percolative model.) The maxima of the two functions agree: both fall at or near optimal doping, $x = x_0 = 0.16$. However, the widths of $T_c(x)$ are larger than those of $f(x)$ on both the underdoped and overdoped sides. This can be the result of dilution of the optimally doped superconductive regions in the unannealed samples by nanoscale phase separation, that is, increased overlapping of insulating and Fermi liquid nanophases with the intermediate phase.

Next, one can use $f(x)$ to identify two ideal (annealed) dopant phase transitions, (1) from the insulating phase to the intermediate superconducting phase at $x = x_1$, and (2) from the latter to the Fermi liquid, at $x = x_2$. Notice that in $f(x)$ the first transition is continuous (second order), while the second transition is first order, as predicted[12] by analogy with the two dopant transitions observed in semiconductor impurity bands. In unannealed (quenched) samples in $T_c(x)$, both the first and the second dopant transitions are greatly broadened into miscibility gaps, $x_1 \to (x_{11}, x_{12})$, and $x_2 \to (x_{21}, x_{22})$. The spinodal tie line of the first gap, $(x_{11}, x_{12})$, is linear, as one would expect from weak dopant-host coupling associated with broadening the second-order transition in $f(x)$. On the other hand, the changes in dielectric screening are large at the first-order transition, $x = x_2$, to the Fermi liquid, just as in impurity bands (Phillips 1999a,b), so it is not surprising that the spinodal tie line of the second gap, $(x_{21}, x_{22})$, is sublinear.

Both for YBCO and for LSCO the *dopant* miscibility gaps are *complementary* to the *host* miscibility gaps; for YBCO this is clear, because the superlattice chain-ordering $(a_0, 2a_0)$ and $(2a_0, 3a_0)$ host gaps are easily identified by diffraction[1]. Then in YBCO the $x = 6.75$ *dopant* miscibility gap must be related to glassy internal coordinates not accessible to diffraction. Because of their specifically complementary orbital character, it is quite obvious that these glassy internal coordinates are not related to spins. In the filamentary model (Phillips and Jung 2001; Phillips 1999a,b) these coordinates are related to self-organized dopant percolation. The dopant $(x_{11}, x_{12})$ and $(x_{21}, x_{22})$ miscibility gaps in Fig. 2 can be explained similarly, without invoking any kind of host structure. This is gratifying, because the only known LSCO secondary host superlattice structure (the 1/8 phase, or "stripes") is known to suppress HTSC.

A second approach to dopant phase diagrams is to fix the primary dopant composition x at its optimal value $x_0$, and to change the composition y of a secondary dopant. It turns out that this can lead to surprisingly simple ("universal") results in some cases. For instance, the isotope effect in x-optimized alloys with secondary dopants, such as $LSr_{0.15}(Cu_{1-y}Ni_y)O$ and $Y(Ba,La)CO_7$ or $(Y,Pr)BCO_7$ is observed (Schneider and Keller 2001) to be linear in $T_c$ over the reduced temperature range $0.3 < T_c/T_m < 0.7$, where $T_m$ is the maximum value of $T_c$. Most surprising here is that these secondary alloys include both LSCO (no host chains) and YBCO (host chains) bases. It was suggested (Schneider and Keller 2001) that in these materials the superconductive layers have decoupled homogeneously and have become nearly two-dimensional (2D) because of their proximity to the critical point for the insulator-superconductor transition to the nearly two-dimensional AF phase, that is, small values of $x - x_1$ or $y - y_1$.

The present model of host and dopant phase diagrams suggests a different explanation. First, consider $LS(Cu_{1-y}Ni_y)O$. The reduced temperature range $0.3 < T_c/T_m < 0.7$ corresponds quite well to the miscibility gap $x_{l1} = 0.06 < x < x_{l2} = 0.12$ in underdoped LSCO alloys (Fig. 2). Because the dopant tie line between $x_{l1}$ and $x_{l2}$ is nearly linear in $T_c$, the isotope shifts $\beta(x)$ should be linear as well. The comparative advantage of this explanation is that it does not have to explain how critical point concepts, usually valid over a range of $\delta x/\delta x_0 = (x - x_1)/(x_0 - x_1) < 0.01$, could be expanded enormously to apply to values of order $(0.12 - 0.05)/(0.16 - 0.05) \sim 0.6$. Moreover, the "universal" agreement with the YBCO-based alloys $Y(Ba,La)CO$ or $(Y,Pr)BCO$ is also easily understood. The reduced temperature range $T_c/T_m < 0.7$ corresponds to $x < 0.60$ in the parent $YBCO_{6+x}$. In this range the host lattice is in the ortho II or $2a_0$ simple alternating chain superstructure. The secondary alloying of $Y(Ba,La)CO$ or $(Y,Pr)BCO$ destroys the $3a_0$ chain superstructure, leaving either the same $2a_0$ superstructure, or none at all, as in LSCO, and a similar miscibility gap, so that the three underdoped secondary phase diagrams can all resemble the LSCO phase diagram of Fig.2, with a similar $(y_{11}, y_{12})$ miscibility gap.

So far the most dramatic and most informative phase diagram for the pseudogap temperature T* appears to have been obtained (Konstantinovic *et al.* 2001) for the one-cuprate plane, no-chain alloy $Bi_2Sr_{1.6}La_{0.4}CuO_y$ (BSLCO). Here oxygen plays the role of the minority dopant, and the majority dopant Sr composition has been tuned to optimize $T_c$. The experimental phase diagram of T*(y) is traced in Fig. 3. (Here y has not been measured directly, so the abcissa used is $\sigma(y)/\sigma(y_{op})$ measured at T = 300K.) The trace represents the present author's interpretation of the experimental points for two sets of samples, which are very close to each other. The trace shows what the author believes to be a clear-cut pseudospinodal tie line, indicative of a first-order transition and a miscibility gap. Even though there are many data points (nine in the linear pseudospinodal region alone), other curves might be traced through the data. However, here the data will be interpreted in terms of the traced curve, including the pseudospinodal region, because this interpretation identifies, in the author's view, some important aspects of the factors contributing to pseudogap formation.

In the upper left corner (a), to the left of the arrow in Fig. 3, the pseudogap can be assigned to an AF spin gap. (This is just what one would expect. The complementary character of the pseudogap and the superconductive gap in the underdoped region is evident from $dT^*/dx < 0 < dT_c/dx$.) This region just barely overlaps the superconductive (SC) region, and it is possible that the AF and SC regions of the sample are spatially disjoint. This region is followed by the miscibility gap, which indicates that in region (b) the pseudogap is probably not a spin gap. Compared to the LSCO phase diagram, Fig. 2, there seems to be good correspondence between the BSLCO miscibility gap and the ($x_{l1}$, $x_{l2}$) LSCO miscibility gap. In LSCO beyond optimal doping T* continues on the linear spinodal tie line until it is close to $T_c$. Here, however, at the optimally doped edge of the miscibility gap, where T* = 200K, there is a steep decrease with small increases in $\sigma(y)/\sigma(y_{op})$, and we can see that the pseudogap near (b) must be different from the spin pseudogap near (a). A possible candidate for this striking behavior is selective depletion of the O occupancy of the "apical" site bonded to both Bi and La, or simply to a Bi – O – Bi site. This is the natural choice for initial reduction, as the Bi – O bonding energy is smaller than the Cu-O bonding energy; it also explains why $T^*(y)$ drops much more steeply in BSLCO than in LSCO, which contains no Bi.

An effect similar to the steep decrease of $T^*$ shown in Fig. 3 for BSLCO is also observed (Segawa and Ando 2001) in detwinned samples of YBCO, especially in the a-axis resistivity, where it is described as "bunching". This bunching occurs exactly in the electronic miscibility gap $(x_{1b}, x_{2b})$ marked in Fig. 2. The effect is smaller in the b-axis resistivity, which again suggests the importance of interchain disordered "defect" bridges between the b-axis chains (Phillips 2001).

In conclusion, there is a good deal of evidence that suggests that the intermediate phase responsible for HTSC is orbital in nature. It is often separated from the AF insulating phase and the Fermi liquid phase by dopant miscibility gaps. These gaps are enlarged by sample quenching or by the presence of stresses that disrupt filaments.

Many experiments have shown that localized spins or virtual spin excitations can coexist with superconductivity in HTSC. To the extent that the orbital miscibility gaps are universal, one can infer that these spin effects are not associated directly with HTSC, but are present, possibly on a nanoscale, as separate dopant phases. Their correlation with $T^*$ further suggests that spin interactions are concentrated in the nanodomain walls, not in the electrically active filaments. In other words, spin-charge separation is a natural result of nanoscale phase separation driven by the formation of ferroelastic nanodomains.

**Figure Captions**

Fig. 1.   Pseudospinodal dopant curves for differently prepared samples of $YBCO_{6+x}$. Curve (a) refers to unstressed, well-annealed samples (Akoshima and Koike 1998), curve (b) to detwinned samples (Segawa and Ando 2001), where the a lattice constant may have been clamped by the detwinning process.  Here and in later figures, the maximum temperatures of the immiscibility domes are not known, and the dotted lines are drawn merely to emphasize the limits of the observed miscibility gaps.  In principle, further information on these domes could be obtained by quenching granular samples with constant grain sizes, and decreasing these sizes between samples, but the detailed nature of the domes is relatively unimportant, once their positions are identified.

Fig. 2.   The Meissner filling factor in $La_{2-x}Sr_xCuO_4$ (LSCO) extremely well annealed samples (Takagi *et al.* 1992).  The first-order character of the strange metal – Fermi liquid transition near x = 0.21 is unambiguous.  Miscibility gaps in $T_c(x)$ [data from (Schneider and Keller 2001) with the schematic lines drawn by the author] correspond well to a diluted optimal phase.

Fig. 3.   The author's trace of the composition dependence of T*(y) in $Bi_2Sr_{1.6}La_{0.4}CuO_y$, (Konstantinovic *et al.* 2001).  In the upper left (a), to the left of the first arrow, the pseudogap is a spin gap.  The region in the lower right, near optimal doping, where T* changes very rapidly while there are only small changes in both $T_c$ and $\sigma(300K)$, is quite mysterious, and is discussed in detail in the text.

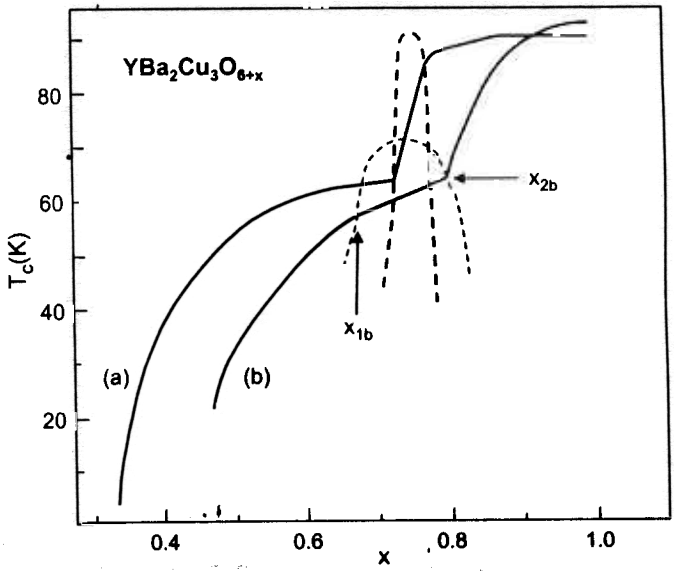

Fig 1

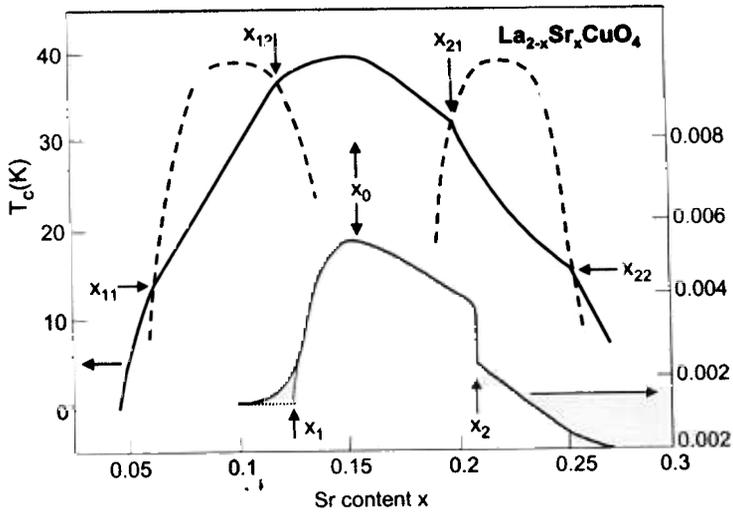

Fig 2.

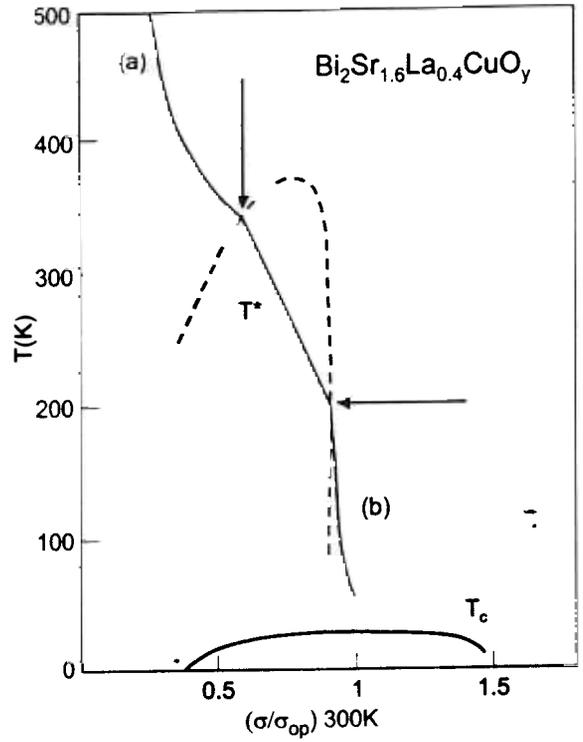

Fig. 3